\documentstyle[twoside,fleqn]{article}
\makeatletter
\def\fileversion{v2.6}
\def\filedate{24 November 1993}

\newdimen\@bls                    
\@bls=\baselineskip               
\advance\@bls -1ex                
\newdimen\@eps                    %
\def\section{\@startsection{section}{1}{\z@}
  {1.5\@bls plus 0.5\@bls}{1\@bls}{\normalsize\bf}}
\def\subsection{\@startsection{subsection}{2}{\z@}
  {1\@bls plus 0.25\@bls}{\@eps}{\normalsize\bf}}
\def\subsubsection{\@startsection{subsubsection}{3}{\z@}
  {1\@bls plus 0.25\@bls}{\@eps}{\normalsize\bf}}
\def\paragraph{\@startsection{paragraph}{4}{\parindent}
  {1\@bls plus 0.25\@bls}{0.5em}{\normalsize\bf}}
\def\subparagraph{\@startsection{subparagraph}{4}{\parindent}
  {1\@bls plus 0.25\@bls}{0.5em}{\normalsize\bf}}

\def\@sect#1#2#3#4#5#6[#7]#8{\ifnum #2>\c@secnumdepth
  \def\@svsec{}\else 
  \refstepcounter{#1}\edef\@svsec{\csname the#1\endcsname.\hskip0.5em}\fi
  \@tempskipa #5\relax
  \ifdim \@tempskipa>\z@
    \begingroup 
      #6\relax
      \@hangfrom{\hskip #3\relax\@svsec}{\interlinepenalty \@M #8\par}%
    \endgroup
    \csname #1mark\endcsname{#7}\addcontentsline
      {toc}{#1}{\ifnum #2>\c@secnumdepth \else
        \protect\numberline{\csname the#1\endcsname}\fi #7}%
  \else
    \def\@svsechd{#6\hskip #3\@svsec #8\csname #1mark\endcsname
      {#7}\addcontentsline{toc}{#1}{\ifnum #2>\c@secnumdepth \else
        \protect\numberline{\csname the#1\endcsname}\fi #7}}%
  \fi \@xsect{#5}}

\long\def\@makefigurecaption#1#2{\vskip 10mm #1. #2\par}

\long\def\@maketablecaption#1#2{\hbox to \hsize{\parbox[t]{\hsize}
  {#1 \\ #2}}\vskip 0.3ex}

\def\fnum@figure{Figure \thefigure}
\def\figure{\let\@makecaption\@makefigurecaption \@float{figure}}
\@namedef{figure*}{\let\@makecaption\@makefigurecaption \@dblfloat{figure}}

\def\table{\let\@makecaption\@maketablecaption \@float{table}}
\@namedef{table*}{\let\@makecaption\@maketablecaption \@dblfloat{table}}

\textfloatsep=\floatsep           
\intextsep=\floatsep              

\long\def\@makefntext#1{\parindent 1em\noindent\hbox{${}^{\@thefnmark}$}#1}

\def\maketitle{\begingroup        
    \def\thefootnote{\fnsymbol{footnote}}%
    \newpage \global\@topnum\z@ 
    \@maketitle \@thanks
  \endgroup
  \let\maketitle\relax \let\@maketitle\relax
  \gdef\@thanks{}\let\thanks\relax
  \gdef\@address{}\gdef\@author{}\gdef\@title{}\let\address\relax}

\def\justify@on{\let\\=\@normalcr
  \leftskip\z@ \@rightskip\z@ \rightskip\@rightskip}

\newbox\fm@box                    

\def\@maketitle{
  \global\setbox\fm@box=\vbox\bgroup
    \vskip 8mm                    
    \raggedright                  
    \hyphenpenalty\@M             
    {\Large \@title \par}         
    \vskip\@bls                   
    {\normalsize                  
     \@author \par}               
    \vskip\@bls                   
    \@address                     
  \egroup
  \twocolumn[
    \unvbox\fm@box                
    \vskip\@bls                   
    \unvbox\abstract@box          
    \vskip 2pc]}                  

\newcounter{address} 
\def\theaddress{\alph{address}}
\def\@makeadmark#1{\hbox{$^{\rm #1}$}}   

\def\address#1{\addressmark\begingroup
  \xdef\@tempa{\theaddress}\let\\=\relax
  \def\protect{\noexpand\protect\noexpand}\xdef\@address{\@address
  \protect\addresstext{\@tempa}{#1}}\endgroup}
\def\@address{}

\def\addressmark{\stepcounter{address}%
  \xdef\@tempb{\theaddress}\@makeadmark{\@tempb}}

\def\addresstext#1#2{\leavevmode \begingroup
  \raggedright \hyphenpenalty\@M \@makeadmark{#1}#2\par \endgroup
  \vskip\@bls}

\newbox\abstract@box              

\def\abstract{%
  \global\setbox\abstract@box=\vbox\bgroup
  \small\rm
  \ignorespaces}
\def\endabstract{\par \egroup}

\def\thebibliography#1{\section*{REFERENCES}\list{\arabic{enumi}.}
  {\settowidth\labelwidth{#1.}\leftmargin=1.67em
   \labelsep\leftmargin \advance\labelsep-\labelwidth
   \itemsep\z@ \parsep\z@
   \usecounter{enumi}}\def\makelabel##1{\rlap{##1}\hss}%
   \def\newblock{\hskip 0.11em plus 0.33em minus -0.07em}
   \sloppy \clubpenalty=4000 \widowpenalty=4000 \sfcode`\.=1000\relax}

\newcount\@tempcntc
\def\@citex[#1]#2{\if@filesw\immediate\write\@auxout{\string\citation{#2}}\fi
  \@tempcnta\z@\@tempcntb\m@ne\def\@citea{}\@cite{\@for\@citeb:=#2\do
    {\@ifundefined
       {b@\@citeb}{\@citeo\@tempcntb\m@ne\@citea
        \def\@citea{,\penalty\@m\ }{\bf ?}\@warning
       {Citation `\@citeb' on page \thepage \space undefined}}%
    {\setbox\z@\hbox{\global\@tempcntc0\csname b@\@citeb\endcsname\relax}%
     \ifnum\@tempcntc=\z@ \@citeo\@tempcntb\m@ne
       \@citea\def\@citea{,\penalty\@m}
       \hbox{\csname b@\@citeb\endcsname}%
     \else
      \advance\@tempcntb\@ne
      \ifnum\@tempcntb=\@tempcntc
      \else\advance\@tempcntb\m@ne\@citeo
      \@tempcnta\@tempcntc\@tempcntb\@tempcntc\fi\fi}}\@citeo}{#1}}

\def\@citeo{\ifnum\@tempcnta>\@tempcntb\else\@citea
  \def\@citea{,\penalty\@m}%
  \ifnum\@tempcnta=\@tempcntb\the\@tempcnta\else
   {\advance\@tempcnta\@ne\ifnum\@tempcnta=\@tempcntb \else
\def\@citea{--}\fi
    \advance\@tempcnta\m@ne\the\@tempcnta\@citea\the\@tempcntb}\fi\fi}

\def\ps@crcplain{\let\@mkboth\@gobbletwo
     \def\@oddhead{\reset@font{\sl\rightmark}\hfil \rm\thepage}%
     \def\@evenhead{\reset@font\rm \thepage\hfil\sl\leftmark}%
     \let\@oddfoot\@empty
     \let\@evenfoot\@oddfoot}

\ps@crcplain                    

\newcommand{\AmS}{{\protect\the\textfont2
  A\kern-.1667em\lower.5ex\hbox{M}\kern-.125emS}}

\makeatother

\newcommand{\bee}{\begin{equation}}
\newcommand{\ene}{\end{equation}}

\newcommand{\al}{\alpha}
\newcommand{\be}{\beta}

\title{1. The counting function.\quad   2. Hybrid boundary conditions.}

\author{{J.S.Dowker}
\address{Department of Theoretical Physics, The University of Manchester,
Manchester, UK. }
\thanks{Supported by EPSRC Grant No. GR/M45726.}}

\begin{document}
\begin{abstract}
\end{abstract}
\maketitle
My talk is divided into two separate parts on the grounds that if you don't
like one then you might like the other. However it leaves little time
for details or references or even accuracy. 

\section{The counting function.}

In any discussion of eigenvalues, the counting function, $N(\lambda)$ is the
basic spectral object. It contains all you want to know, globally speaking.

Probably the first example of such a function occurs in Sturm's 1829
treatment of the roots of the secular equation. Later in the century it
arose in connection with cavity black body radiation (Rayleigh and Jeans)
when the importance of the asymptotic behaviour of $N(\lambda)$ for large 
$\lambda$
became physically more pressing.

Weyl showed that, for the Laplacian in a cavity of volume $V$,
$
N(\lambda),\equiv\sum_{\lambda_m<\lambda} 1$, has the asymptotic form,
\bee
N(\lambda)\sim{V\over(4\pi)^{3/2}} {\lambda^{3/2}\over \Gamma(3/2+1)}, \quad 
{\rm as}\quad\lambda\to\infty. \label{weyln}
\ene
It was natural to ask about corrections to this leading
form. Unfortunately $N(\lambda)$ is an awkward quantity to work with. It's not
smooth. Recall the term `spectral staircase'. 

To handle this one takes various {\it transforms}, for example,
the Laplace transform giving the heat-kernel, $K(t)$,
\bee
K(t)=\int_{0^-}^\infty e^{-\lambda t}\, dN(\lambda)\,, \label{countform1}
\ene from
which spectral information can be extracted, looking at it as a function of
$t$. $K$ is much easier to deal with than $N$ and so maybe one can find
out about large $\lambda$ $N$ from small $t$ $K$ by inversion.
This is the Carleman approach. Sounds easy.

Under certain fairly wide conditions, $K(t)$
has a small-$t$ expansion (a `pre-Tauberian result')
\bee K(t)\sim\sum_{n=0}^\infty
c_{\al_n}\,t^{-\al_n}\,,\quad t\downarrow0, \label{genasymp}\ene
where the indices $\al_n$ start off positive ($\al_0=3/2$) and decrease 
to minus infinity. The coefficients, $c_{\al_n}$, are usually locally
computable and often have a geometrical significance.

By Laplace inversion, the asymptotic form for a `smoothed' counting
function,
\bee \overline
 N(\lambda)\sim\sum_{n=0}^\infty 
{c_{\al_n}\over\Gamma(1+\al_n)}\,\lambda^{\al_n}\,,
 \label{smasymp}\ene
follows as a `post-Tauberian result'. I note that there are 
{\it no terms with $\al_n$ a negative integer}. 

This result is due to Brownell who pointed out the
need for smoothing in order to pick up the non-leading terms in the
expansion.

The oscillating difference, $N(\lambda)-\overline N(\lambda)$, is the concern
of Periodic Orbit Theory and forms the subject of Fulling's talk here.

Just a few words on a rough and ready justification of (4). The relation is
trivially true for $\al_n$ greater than $-1$ by virtue of Euler's Gamma
function formula,
\bee{1\over t^\al}={1\over\Gamma(\al)}\int_0^\infty
 \lambda^{\al-1}e^{-\lambda t}\,d\lambda,\,\, t>0\,,{\rm Re\, } 
\al>0\,. 
\label{egamma}
\ene For
other $\al_n$ there is a divergence at the lower $\lambda$ integration 
limit in (2) which
can be circumvented by differentiating with respect to $t$ a sufficient number
of times, and then
integrating back, a device used by Fulling (Siam.J.Math.Anal. {\bf13} (1982) 
891) and also by van der
Pol and Bremmer in their book ``Operational Calculus'', 1955.
You can see that the terms in the expansion of $K(t)$, (3), with
positive integer powers ($\al_n$ a negative integer) arise during
this reverse process via constants of integration and are divorced from the 
terms evident in (4) ({\it cf} Heaviside and Carson in the context of
circuit theory). 

As I said, such terms are absent from the asymptotic, large $\lambda$ form of 
the
smoothed counting function $\overline N(\lambda)$, (4). However there is a 
relic of
them as delta-function derivatives, as noted by Van der Pol and Bremmer.
Saying it another way,
these terms {\it can} appear as genuine large $\lambda$ terms, but only after
integrating $\overline N(\lambda)$ a sufficient number of times. 
These integrations, or {\it further smoothings}, 
are best performed using {\it Riesz means}, as discussed systematically by
Fulling and Estrada (Elec.J.Diff.Equns.{\bf1999} No.6,1; No.7,1)
and this leads me onto what I want to say here. 

As 
initiation, I go back to the general expansion of the smoothed $\overline
N(\lambda)$, (4) and give it a touch with the distributional 
wand by interpreting it as,
 \bee
\overline N(\lambda)=\sum_{n=0}^\infty c_{\al_n} \Phi_{\al_n+1}(\lambda),
 \label{distn}\ene
where the {\it generalised function}, $\Phi_\al(x)$, is
 \bee
\Phi_\al(x)={x^{\al-1}_+\over\Gamma(\al)}\,.
 \label{genp}\ene
I remind you that the generalised function $x^\al_+$ is concentrated
on the positive $x$-axis, {\it i.e.}
$x^\al_+$ is equal to $x^\al$ for $x\ge0$ and is zero for $x<0$. See
your favourite book on generalised functions -- {\it e.g.} Gelfand and Shilov.
$\Phi_\al$ has many nice properties as a function of $\al$.

The Laplace transform of $\overline N(\lambda)$ now yields {\it all} the small
$t$ expansion of $K(t)$, (3), by virtue of the fact that,
 $$
\int_0^\infty \Phi_\al(x) e^{-t x}\,dx={1\over t^\al}\,,\quad t>0\,,
 $$
is valid for {\it all} $\al$ by the uniqueness of analytic continuation (
Gelfand and Shilov, Oldham). (This is the Gamma function formula again.)

One of the nice properties of $\Phi_\al$ is,
 \bee
\Phi_{-k}(x)=\partial_x^k\delta(x)\,,\quad k=0,1,2,\ldots\,,
 \label{deriv}\ene
which corresponds to those terms in the series for $K(t)$ with positive 
integral powers of $t$ that I mentioned earlier.

Moreover, by making use of the notion of {\it fractional derivative}, 
the restriction
to integral negative $\al_n$ can be removed and one has the compact
expression,
 \bee
\overline N(\lambda)\sim\sum_{n=0}^\infty c_{\al_n}\partial_\lambda^{-\al_n-1}
\delta(\lambda)\,,
 \label{deltn2}\ene
for  {\it all} $\al_n$,
and this is my final equation relating the heat--kernel and counting function
series.

In this way one can give a unified, attractive (if dangerous) Cauchyesque
treatment of the {\it whole} series based on the convolution property of the
function $\Phi_\al$,
 $$ \Phi_\al*\Phi_\be=\Phi_{\al+\be}\,.
 $$
\section {\bf The hybrid spectral problem.}

Again I am interested in the eigenvalue problem in some domain 
but this time the setting
is a little non-standard. In particular, the eigenfunctions satisfy Dirichlet
on one part of the boundary and Neumann on the rest. We will hear more
about this from Avramidi, Gilkey and Seeley. I will not discuss the
problem in any generality but just begin with some examples 
(see hep-th/0007129)
which, no doubt, fall within the framework set up by Br\"uning and Seeley, 
J.Func.Anal. {\bf95} (1991) 255, who refer to
the Zaremba problem. (This is a generalisation of the Dirichlet potential 
problem. See Sneddon, ``Mixed boundary value problems'' North-Holland, 1966, 
and the history in Azzam and Kreysig, Siam.J.Math.Anal. {\bf13} (1982) 
254-262.)

I start with the simplest situation -- the Laplacian on the interval,
$[0,L]$, with $D$ and $N$ conditions at the ends.

A simple calculation or (better) the drawing of a few low modes, shows that 
the various eigenproblems are functorially related by,
\begin{eqnarray}
(D,N)_L\cup(D,D)_L&=(D,D)_{2L}\\ (D,N)_L\cup(N,N)_L&=(N,N)_{2L}
\label{relns}
\end{eqnarray}
\bee (D,D)_L\cup(N,N)_L=P_{2L}\,, \label{relns2}\ene 
where the
notation $(D,N)$ signifies a problem with $D$ conditions at one end and $N$
at the other. $P$ stands for periodic conditions.

The `subtraction' implied by (10), in order to extract the $(D,N)$
part, amounts to a cull of the even modes on the doubled interval, as is
well known ({\it cf} Rayleigh ``Theory of Sound", 1894 vol. I, p.247).

Averaging (10) and (11) and using (12) yields an expression for $(D,N)$
purely in terms of periodic quantities,
\bee
(D,N)_L\cup{1\over2}P_L={1\over2}P_{2L}.
\ene

The relations for rectangular regions like $[0,L]\times[0,L]$ can be 
obtained by formal multiplication of (10)-(12). I will not consider them here
but pass on to 
\vskip5truept
\noindent{\bf The wedge.}
\vskip5truept
The interval relations can be applied to the arc of a circle, which 
might form
part of an SO(2) foliation of a two--dimensional region (or the projection
of a higher dimensional region onto two dimensions). A wedge is a good
example which I will now consider. Let the angle of the wedge be $\be$,
then the relations (10) (11) apply with $L\equiv\beta$, where the notation
now means that either $D$ or $N$ holds on the straight sides of the wedge,
(say $\phi=0$ and $\phi=\be$).

The relations (10) (11) can be immediately
applied to the heat-kernel and its small-time expansion to determine the
form of the heat-kernel coefficients in the $(D,N)$ combination. I will
show how this works out for the $C_1$ heat-kernel coefficient in two 
dimensions. For the 
Laplacian, the $(D,D)$ and $(N,N)$ wedge coefficients are well known,
\bee C^{\rm wedge}_1(D,D)=C^{\rm wedge}_1(N,N)=(\pi^2-\be^2)/6\be.
\label{cee1s}\ene 
Hence, from (10) or (11), algebraic subtraction gives, \bee C^{\rm
wedge}_1(D,N)=-(\pi^2+2\be^2)/12\be. \label{cee1DN}\ene This result has
been derived by Simon Watson (Thesis, University of Bristol 1998) 
using the Bessel function modes directly. Clearly, when $\be=\pi$ 
(no physical wedge), there is still an effect. The $N\cap D$ join, 
$\Sigma$, is a singular region.

Sommerfeld, in Frank u.von Mises' ``Differential und Integralgleichungen'' 
1930, vol.2 p.827, also mentions the `mixed'
wedge and indicates how to treat it using images if $\be=\pi n/m$ and refers
to hydrodynamical applications by Zeilon (1924) of the potential equation
({\it i.e.} the Zaremba problem).
\vskip5truept
\noindent{\bf The $C_1$ coefficient. A conjecture.}
\vskip5truept
Heuristic arguments suggest the following general form for the $C_1$
heat-kernel coefficient in the $R/D$ case, ($R$ = Robin),
\begin{eqnarray}&C_1(D,R)=\big({1\over6}-\xi\big)\int Rf dV+\\
&\!\!\!\int_D\!\big({\kappa\over3}\!
-\!{1\over2}(n.\partial)\!\big)fdS+\!\int_R\!\big({\kappa\over3}\!
-\!2\psi\!+\!{1\over2}(n.\partial)\!\big)fdS\nonumber\\
&\!\!+\int_{(D,D)
\cup(R,R)} {\pi^2-\be^2\over6\be}\,fdL-\int_{(D,R)}{\pi^2+2\be^2\over12\be}
fdL. \nonumber\label{cee1}\end{eqnarray}
$\psi$ is the Robin function and dimensions show that it cannot enter into
the codimension--2 contribution (the final two integrals).
\vskip5truept
\noindent{\bf The 2--lune.}
\vskip5truept
The next simplest geometry (of constant curvature) is a lune, that is, a 
sector of a sphere.

The intervals to which relations (10) (11) are applied 
are the sections of the lines of latitude cut out by the two
longitudes, $\phi=0$, $\phi=\be$. The singular region, $\Sigma$, comprises 
the two poles which, in higher dimensions, are really full spheres.

Denoting the lune by ${\cal L}(\be)$ we have exactly the same relations,
\begin{eqnarray}
(D,N)_{{\cal L}(\be)}\cup(D,D)_{{\cal L}(\be)}&=(D,D)_{{\cal L}(2\be)}
\nonumber\\
(D,N)_{{\cal L}(\be)}\cup(N,N)_{{\cal L}(\be)}&=(N,N)_{{\cal L}(2\be)}\,,
\end{eqnarray} so that
the corresponding zeta--functions\ combine algebraically, 
\begin{eqnarray}
\zeta_\be^{ND}(s)&=\zeta_{2\be}^{DD}(s)-\zeta_\be^{DD}(s)\nonumber\\
&=\zeta_{2\be}^{NN}(s)-\zeta_\be^{NN}(s)\,. \label{ndzeta}\end{eqnarray}

The $DD$ and $NN$ zeta--functions\ are given in terms of Barnes 
zeta--functions. They can be used
to confirm the expression (16) using the
relation between $C_1$ and $\zeta(0)$ in two dimensions. 
In this case the {\it extrinsic}
curvatures, $\kappa$, vanish (the boundaries are geodesically embedded) 
but there is a
volume (area) term independent of the boundary conditions. Higher dimensional
spheres can be treated easily.
\vskip5truept
\noindent{\bf The 2--hemisphere and disc.}
\vskip5truept
Setting $\be=\pi$ gives the hemisphere which is conformally related to the
flat {\it disc}.

Now, the disc with $N$ on one half of the
circumference and $D$ on the other seems hard to treat, in so
far as the construction of the heat-kernel or zeta-function is concerned. 
However, granted the general form of the $C_1$ coefficient, (16), 
it is possible to determine the
{\it functional determinant} for conformally invariant fields on the $N/D$
disc from that on the hemisphere, which can be found from the Barnes zeta
function and is, up to exponentiation,
\bee {\zeta^{ND}_{\pi}}'(0)=-\zeta_R'(-1)-{1\over12}\log2-{1\over4}.
\label{NDdet}\ene

For comparison, the standard formulae for the $DD$ and $NN$-hemispheres are,
$$ {\zeta^{DD}_\pi}'(0)=2\zeta_R'(-1)-\zeta_R'(0)-1/4 $$ and $$
{\zeta^{NN}_\pi}'(0)=2\zeta_R'(-1)+\zeta_R'(0)-1/4. $$

By stereographic transformation to the $N/D$--disc, our final result 
turns out to be, $$
{\zeta_{ND}^{\rm disc}}'(0)=-\zeta_R'(-1)-{11\over12}\log2+{1\over12}\,. $$ 

\noindent{\bf The $d$-dimensional hemisphere/lune.}
\vskip5truept
The zeta--functions are again of Barnes type and the known pole structure 
allows one to extract
the contribution of the singular region, $\Sigma$, to the $C_n$ heat--kernel
coefficients as explicit polynomials in $d$.
\vskip5truept
\noindent{\bf Added notes.}
\vskip5truept
The standard $(D,D)$ and $(N,N)$ results for the wedge, (14), (going back to
Sommerfeld, 1898, and Carslaw, 1910) are a consequence of the, often 
unspoken, use of the Friedrichs extension of the compact Laplacian, which, 
for $\be\le\pi$, is the
{\it only} extension. For $\be>\pi$ there exist ``unphysical'' extensions 
corresponding to the inclusion of $L^2$ radial modes, $J_{-|\nu|}(kr)$ 
with $|\nu|<1$, ({\it e.g.} Cheeger and Taylor, Comm. Pure. Applied 
Math. {\bf35} (1982) p.305; Strichartz, J.Func.Anal. {\bf91} (1990) 37).

The relation between the $(D,N)$ and $(D,D)$ wedge problems, say, shows that
unphysical extensions become possible for the $(D,N)$ case if $\beta>\pi/2$.
This excludes rectangular domains (dropping the $N_n$ solutions)
but for the $\be=\pi$, geometrically smooth case,
an extra mode, $J_{-1/2}$, exists ({\it cf} Seeley's contribution to this 
conference). For $\be>3\pi/2$, {\it two} extra modes appear,
and so on. This shows that, in this regard, there is nothing special about
this $N/D$ problem, and all those others obtained by using the 
relations (10) (11) and, as remarked by Fulling at this conference, the 
freedom of extension does not seem immediately to provide a means of avoiding
the difficulty with the index encountered in Dowker, Kirsten and 
Gilkey (hep-th/0010199). 

Relations (10) (11) also show that there are no $\log t$ terms in the $N/D$ 
heat--kernel expansion if there are none in the {\it related} $DD$ or $NN$ 
cases. Of course this only applies to a very special class of $N/D$ 
situations. In Dowker, Kirsten and Gilkey, an $N/D$ spherical cap 
problem, not obviously covered by (10) (11) but with $\Sigma$ embedded 
smoothly in the  boundary, is discussed perturbatively and suggests, 
albeit indirectly, that $\log t$ terms are present, generated by a non-zero 
extrinsic curvature. This result appears to disagree with the general
conclusion of Seeley presented at this conference. However a detailed check
of the geometry has still to be made in order to
confirm that the example actually does fall into the class covered by 
Br\"uning and Seeley, although Seeley's value for $C_1$  agrees with (15), 
at $\be=\pi$, (which is the same as  the value found by Avramidi).

A point to bear in mind is 
the relation between the $N/D$ and periodic problems on the interval, equation
(13), which shows that, even for the $\be=\pi$, $N/D$ wedge, there is an
effective $4\pi$ conical singularity at $\Sigma$ and conical singularities
generically lead to $\log t$ terms, as shown by Cheeger, (J.Diff.Geom.
 {\bf18} (1983) 575) and by Br\"uning and Seeley, 
(J.Func.Anal. {\bf73} (1987) 369), but not for the wedge.

\end{document}